\documentclass[aps,prb,preprint]{revtex4}
\usepackage{graphicx}

\begin{document}

\title{Ab initio study of single molecular transistor modulated by gate-bias}
\author{F. Jiang}
\affiliation{Department of Physics, Fudan University, Shanghai
200433, People's Republic of China}
\author{Y.X. Zhou}
\affiliation{Department of Physics, Fudan University, Shanghai
200433, People's Republic of China}
\author{H. Chen }
\email[Corresponding author. Email: ] {haochen@fudan.edu.cn}
\affiliation{Department of Physics, Fudan University, Shanghai
200433, People's Republic of China}
\author{R. Note}
\affiliation{Institute for Materials Research, Tohoku University,
Sendai 980-8577, Japan}
\author{H. Mizuseki} \affiliation{Institute for Materials Research, Tohoku
University, Sendai 980-8577, Japan}
\author{Y. Kawazoe}
\affiliation{Institute for Materials Research, Tohoku University,
Sendai 980-8577, Japan}

\date{\today}

\begin{abstract}
 We use a self-consistent method to study the current of the single molecular transistor
 modulated by the transverse gate-bias in the level of the first-principles calculations.
 The numerical results show that both the polyacene-dithiol molecules and the
 fused-ring oligothiophene molecules are the potential high-frequency molecular
 transistor controlled by the transverse field.
 The long molecules of the polyacene-dithiol or
 the fused-ring thiophene are in favor of realizing the gate-bias controlled
 molecular transistor. The theoretical results suggest the related
 experiments.
\end{abstract}

\maketitle
\section{\label{sec:level1}INTRODUCTION}
 Using organic molecules as functional units for electronic
 apparatus application is an interesting goal of nanoelectronic
 devices.\cite{1,2} The common and important function of these devices
 is that the current can be controlled effectively. In the last
 several years, many experimental and theoretical works were
 carried out to study the transport properties through a single
 molecule, or even to design the molecular electronic devices.
 \cite{3,4,5,6,7,8,9,10,11,12,13,14,15,16,17,18,19,20,21,22,23,24,25,26,27,28}
At present, people have realized
 two major approaches to control molecular transport. One
 is through the conformational change in the molecule,
 the other is through the external transverse field to switch the molecule
 from an \textquoteleft\textquoteleft on\textquoteright\textquoteright
 \  to an \textquoteleft\textquoteleft off\textquoteright\textquoteright
 \  state.
 Many authors have focused their attention on the former one for a long
 time.\cite{20,21,22,23,24,25}
 Despite the conformational change in the molecule can be achieved by using
 the electric or light fields, its operational frequency is low. Now, the attention is
 transferred to the latter one due to its high operation frequency. Several
 experiments have identified their feasibility.\cite{26,27,28} Currently,
 the $\pi$-conjugated organic oligomers and polymers are the
 subject of considerable research interest in the organic semiconductors.
 The organic semiconductors can be employed as active
 layers in the field effect transistors (FET).\cite{26}
 The gate-bias controlled molecular transistor is successfully
 achieved experimentally \cite{28}
 from perylene tetracarboxylic diimide (PTCDI), a redox molecule.
 Recently, the current behavior of the single molecule has received increasing attention.
 Single organic oligomers such as pentathienoacene (PTA), pentacene, perylene
 and so on are all the key objects in the theoretical study. But at present,
 few theoretical work sheds light on the organic molecular transistor
 controlled by the transverse field. The rigorous treatment
  of the molecular device in theory calls for the combination of
  the theory of quantum transport with the first-principles calculations
  of the electronic structure in the self-consistent scheme. In this paper,
  we use the density functional theory (DFT) and non-equilibrium Green's
  function to study the transverse field effect (TFE) on current
  transport of the single organic oligomer.

\section{\label{sec:level1}THEORY AND METHODS}
  The retarded Green's function of the molecule is expressed as follows
\begin{equation}G^{R}_{M}=(E^{+}S_{M}-F_{M}-\Sigma^{R}_{1}-\Sigma^{R}_{2})^{-1},
\end{equation}
where $S_{M}$ and $F_{M}$ are the overlap matrix and Fock matrix
of the molecule part, respectively. $\Sigma^{R}_{1}$
($\Sigma^{R}_{2}$) is the retarded self-energy of the left (right)
electrode. It should be emphasized that the Fock matrix is
obtained after the density matrix is obtained.

The density matrix of the open system is the essential function of
the whole self-consistent scheme. It can be achieved by the
Keldysh Green's function
\begin{equation}\rho=\int^{\infty}_{-\infty} dE[-iG^{<}(E)/2\pi],\end{equation}
\begin{equation}-iG^{<}=G^{R}_{M}(f_{1}\Gamma_{1}+f_{2}\Gamma_{2})G^{A}_{M},
\end{equation}
with the advanced Green's function $G^{A}=(G^{R})^{\dagger}$, the
broadening function of the left (right) lead $\Gamma_{1}$
($\Gamma_{2}$). The Fermi distribution function of the left
(right) lead $f_{1}$ ($f_{2}$) is expressed
$f_{i}(E)=1/(e^{(E-\mu_{i})/kT}+1)$ with
$\mu_{1}=\mbox{$E_{f}$}-\frac{1}{2}eV$,
$\mu_{2}=\mbox{$E_{f}$}+\frac{1}{2}eV$. $E_{f}$ is Fermi level of
the bulk Au. In our work, $E_{f}$ is -5.1 eV which is an adjusted
parameter around its work function (5.31 eV) for explaining
experimental results.\cite{30,31}

If the transverse field is applied perpendicular to the transport
direction, the corresponding potential energy term is included in
the Fock operator
\begin{equation}V_{E_{\perp}}(\vec{r})=e\vec{E}_{\perp}\cdot\vec{r}.
\end{equation}

Our model is illustrated in Fig. 1 with the molecule attached by
the gold electrodes from the both sides and the transverse
electric field confined inside the molecule region. The potential
zero point is set at the coordinate origin, the center of the line
connecting two sulfur atoms and the whole molecule is not
symmetric to the Au-S bond. Since the transverse field is
localized in the molecule, its edge effect is very weak so that it
almost has no effect on the electrodes. In the calculation, the
applied longitudinal voltage points to the current direction shown
in Fig. 1. The molecule is chemisorbed on the gold contacts by
sulfur atoms. The sulfur atom sits on the hollow position of three
nearest-neighbor surface gold atoms. The perpendicular distance
between the sulfur atom and the gold FCC (111) surface plane is
2.0 \AA, an usually acceptable distance. The temperature effect is
not distinct for the short molecule, so we assume zero temperature
in our calculation for simplicity.

\begin{figure}
\includegraphics[scale=0.6, angle=0, bb=15 372 557 757]{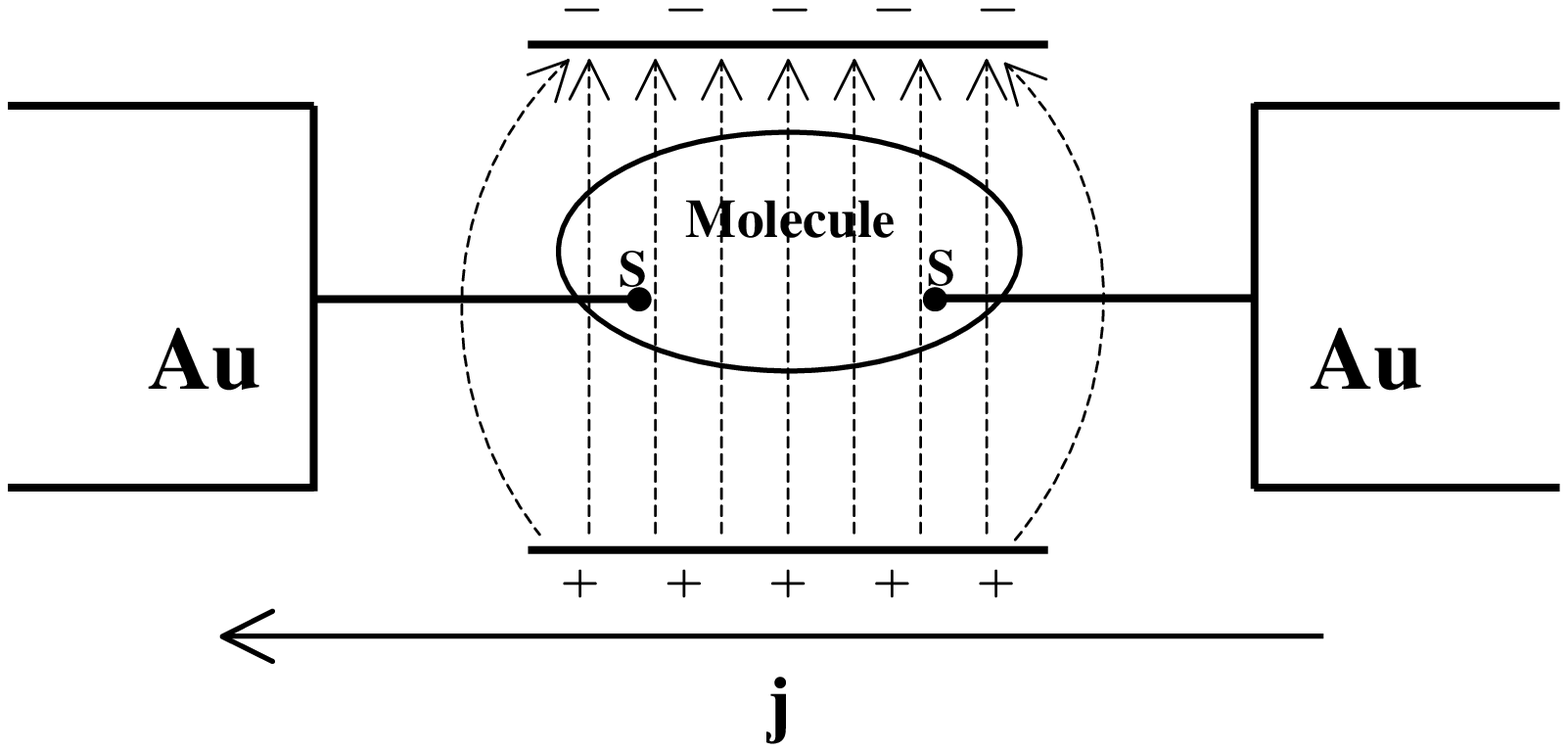}
\caption{Au lead-Molecule-Au lead open system. The molecule is
chemisorbed to Au leads by sulfur atoms and is not symmetric to
the Au-S bond. This transverse field is confined inside the
molecule region without effect on Au leads. \label{fig1}}
\end{figure}

The density matrix for the open system should be obtained
self-consistently. In order to achieve this goal, we extended the
inner loop in the standard quantum chemistry software GAUSSIAN03
\cite{31} to the loop composed of the lead-molecule-lead open
system under bias. At the beginning, the self-consistent procedure
starts from a guess for the density matrix of the open system,
which may be obtained from the converged density matrix of
GAUSSIAN03 calculation for the isolated molecule. We feedback the
density matrix to the GAUSSIAN's main program as a subroutine to
obtain the new density matrix. The iterations continue until the
density matrix converges to the acceptable accuracy (usually less
than $10^{-5}$). Then the density matrix is used to evaluate the
transmission function (T), the terminal current, the electron
number and the density of states (DOS) of the open system.
\cite{29,30} In the calculations, we adopt DFT with B3PW91
exchange-correlation potential and LANL2DZ basis to evaluate the
electronic structure and the Fock matrix. The basis set associates
with the effective core potential (ECP), which are specially
suited for the fifth-row (Cs-Au) elements including the Darwin
relativistic effect.

\section{\label{sec:level1}RESULTS AND DISCUSSION}
The sulfur atom in thiophene is sp$^{2}$ hybridized, and its
p-orbital provides two electrons to the $\pi$ system. The linearly
condensed thiophene molecules possess the extended $\pi$
conjugation and the high planarity. Of the cyclic single organic
molecules, the thiophene-based compound is one of the promising
class of organic materials. PTA is attractive, due to the
stability of the thiophene ring and the good planarity.\cite{26}
Fig. 2 gives the DOS and T of PTA with the different transverse
gate biases. The position of the broadened levels in equilibrium
for the open system is determined by the singular points of the
Green's function, obtained from equation
  $(F+\Sigma_1+\Sigma_2)C=SC\lambda$.
The molecular levels are modulated by the gate bias applied to the
molecule with HOMO (LUMO) position at $-6.47$ ($-3.93$) eV for
$V_g=-3.90$ V (a); $-6.27$ ($-3.77$) eV for $V_g=-1.95$ V (b);
$-6.09$ ($-3.60$) eV for $V_g=0$ V (c); $-5.93$ ($-3.42$) eV for
$V_g=+1.95$ V (d), and $-5.79$ ($-3.24$) eV for $V_g=+3.90$ V (e).
With decrease of the positive gate bias, the separation between
HOMO and Fermi level increases, and the separation between LUMO
and Fermi level decreases. For the positive gate bias $V_g\geq 0$,
the Fermi level is close to HOMO, and PTA is the p-type or hole
conduction molecule, while for the big negative bias PTA is the
n-type or electron conduction molecule. For large positive bias
$V_g=+3.90$ V, the HOMO, close to $E_{f}$, contributes to the
initial rise of the current under the small applied voltage (1.0 V
or so) in the longitudinal direction. Meanwhile LUMO is close to
$E_{f}$ in the case of large negative bias $V_g=-3.90$ V, it is
responsible for the molecular conduction (Fig. 3). Fig. 2
illustrates that In Fig. 3 the gate bias successfully modulates
the I-V characteristics of the PTA. At the voltage $V>-2.3$ V, due
to the contribution from HOMO, the positive gate bias achieves the
bigger molecular current and at the voltage $V<-2.3$ V, as LUMO
enters the voltage window and contributes the molecular
conduction, the inverse order of current appears. The inset
illustrates
  the $\alpha$ electron number deviation from the equilibrium state
  as a function of bias.
  At applied voltage $V>-2.3$ V, for gate bias $V_g=-3.90$ V,
  the electron is responsible for the conduction ($N>N_0$)
  and electron number rises slowly, while for $V_g\ge -1.95$ V,
  the hole is responsible for the conduction ($N<N_0$),
  so electron number descends (the electron flows out of the molecule
  to the lead).
  At applied voltage $V<-2.3$ V,
  the molecule enters the electron- and hole- hybrid conduction region,
  companying the rapid rise of electron number, which means the electron
  contribution dominates the conduction.

\begin{figure}
\includegraphics[scale=0.5,angle=0,bb=21 20 575 819]{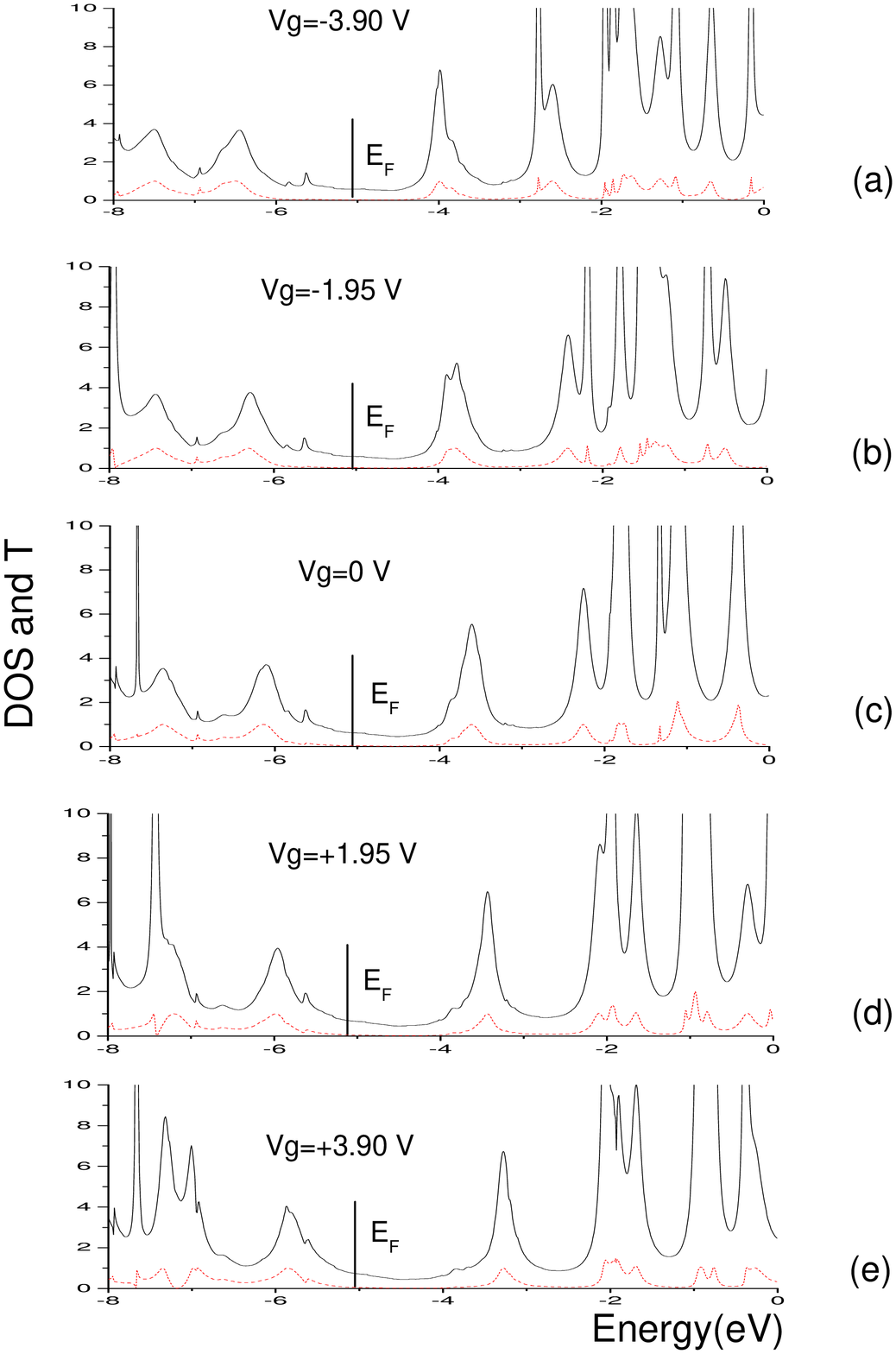}
\caption{DOS (solid) and T (dashed) as functions of energy for PTA
with gold contacts connected, with the transverse gate bias:
$V_g=-3.90$ V (a), $-1.95$ V (b), 0 V (c), +1.95 V (d), and +3.90
V (e). The vertical line denotes the position of Fermi level.
\label{fig2}}
\end{figure}

\begin{figure}
\includegraphics[scale=0.5,angle=-90,bb=42 77 581 788]{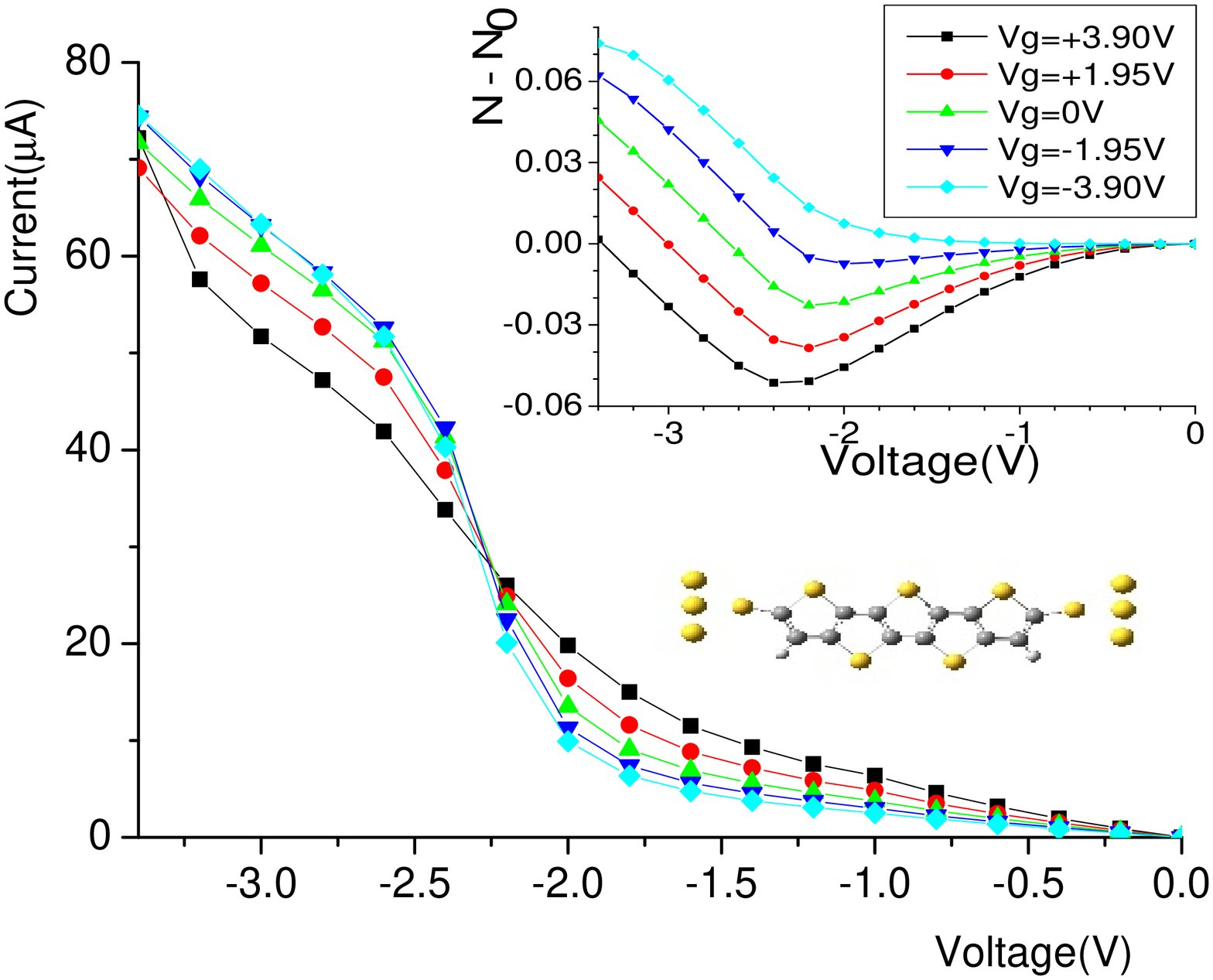}
\caption{Gate modulation of the I-V characteristics for PTA. The
voltage $-2.3$ V is the crossover point for the current controlled
by the gate bias. For voltage $V>-2.3$ V, either electron
($V_g=-3.90$ V) or hole ($V_g\ge -1.95$ V) is responsible for the
current. For voltage $-3.4$ V$<V<-2.3$ V, the molecule enters the
electron- and hole- hybrid conduction region, and the positive
biased current is smaller than the negative ones. \label{fig3}}
\end{figure}

\begin{figure}
\includegraphics[scale=0.5,angle=0,bb=17 15 571 814]{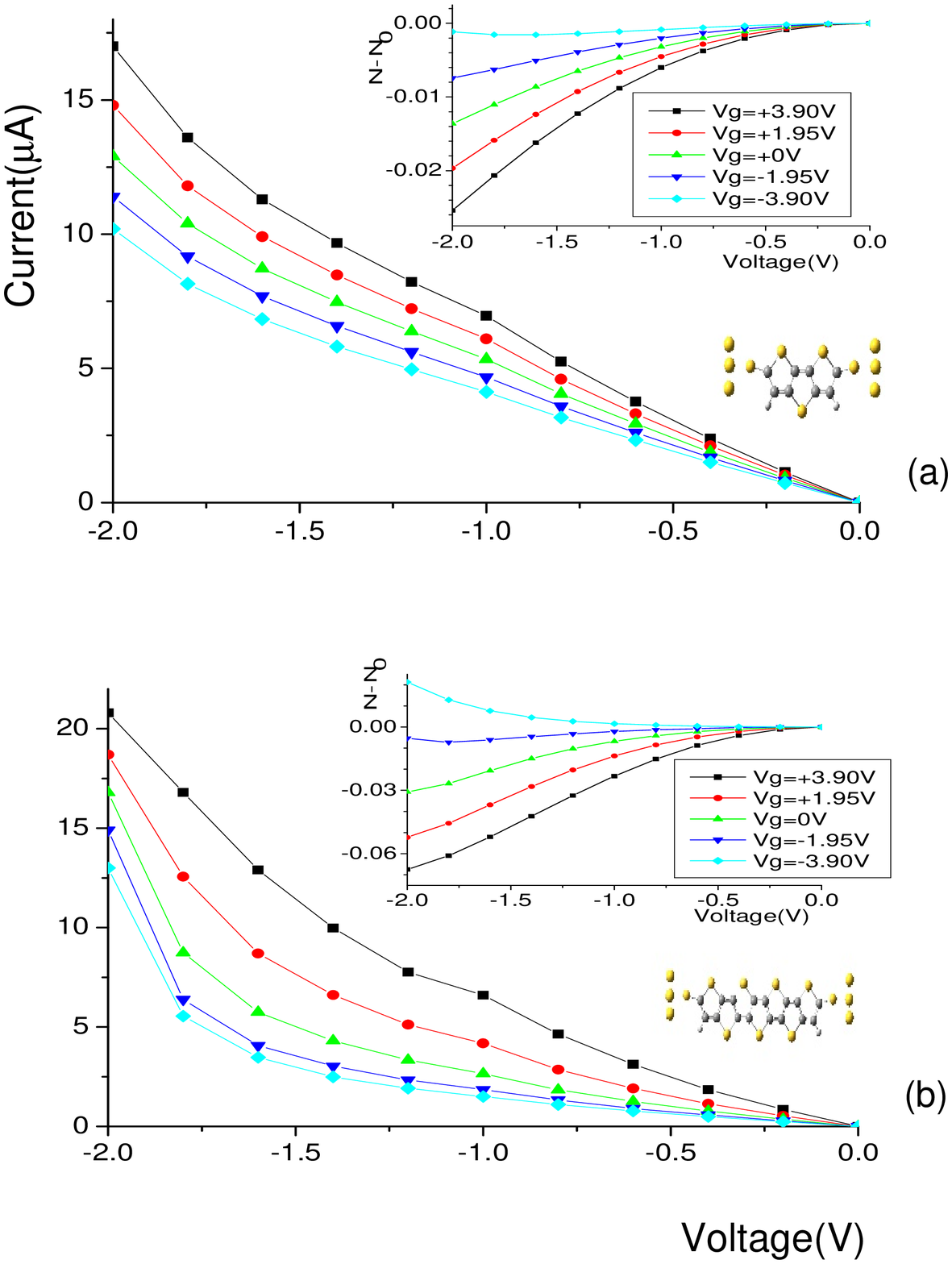}
\caption{I-V curves of two fused-ring thiophenes corresponding to
the different transverse field: 3-fused-ring thiophene (a),
7-fused-ring thiophene (b). The inset shows the $\alpha$ electron
number deviation from the equilibrium state with the modulation of
bias. \label{fig4}}
\end{figure}

Similar to PTA, the good TFE is achieved by the other fused-ring
thiophene molecules. Fig. 4(a) and (b) show the I-V curves of the
3-fused-ring thiophene and 7-fused-ring thiophene controlled by
the gate bias. For example, at applied voltage $ V=-1.0$ V, for
the 3-fused-ring thiophene molecules,
  $(\mbox{I}_{V_g=+1.95 V}-\mbox{I}_{V_g=0 })/(\mbox{I}_{V_g=0 })\doteq 0.14$,
  $(\mbox{I}_{V_g=+3.90 V}-\mbox{I}_{V_g=0 })/(\mbox{I}_{V_g=0 })\doteq 0.30$;
  for the 7-fused-ring thiophene,
  $(\mbox{I}_{V_g=+1.95 V}-\mbox{I}_{V_g=0 })/(\mbox{I}_{V_g=0 })\doteq 0.58$,
  $(\mbox{I}_{V_g=+3.90 V}-\mbox{I}_{V_g=0 })/(\mbox{I}_{V_g=0 })\doteq 1.49$.
Meanwhile at the same bias for PTA, $(I_{V_g=+1.95 V}-I_{V_g=0
})/I_{V_g=0 }\doteq 0.29$, $(\mbox{I}_{V_g=+3.90
V}-\mbox{I}_{V_g=0 })/\mbox{I}_{V_g=0 }\doteq 0.70$. The longer
fused-ring thiophene molecules have better TFE property.

  Although the early experiment \cite{32} reported only weak gate
  effect on the molecular current observed for the single
  benzene-dithiol molecule. Our calculation found that the
  TFE can be improved by the longer polyacene-dithiol molecules,
  which present a good gate-bias controlled molecular transistor property.
  DOS and T of the single PDT, naphthalene-dithiol,
  and anthracene-dithiol in equilibrium without the longitudinal
  and transverse field are shown in Fig. 5, where both HOMO and LUMO are close to
  $E_{f}$ with increase of the aromatic phenyl ring.
  The HOMO (LUMO) is $-7.48$ ($-2.75$) eV for the single PDT,  $-7.44$
  ($-3.55$) eV for the naphthalene-dithiol, and $-6.98$ ($-3.86$) eV
  for the anthracene-dithiol.
  The fact predicts that if the number of aromatic phenyl ring is
  increased to more than 3, HOMO and LUMO will be closer to $E_{f}$, and the
  TFE will be more apparent. For the pentacene, $|\mbox{HOMO}-\mbox{LUMO}|\doteq$ 2 eV and
  $E_{f}$ is almost in the middle of H-L gap, it has apparent TFE in the small bias.

  The anthracene-dithiol is the n-type conduction
  molecule in the small voltage region,
  its LUMO is closer to $E_{f}$ than HOMO [see Fig. 5(c)]. Its
  molecular levels can be modulated obviously by the gate-bias.
  With the gate-bias increased by increment $2.5$ V from $V_g=-5.00$ V to $V_g=+5.00$ V,
  the HOMO of anthracene-dithiol leaves $E_f$ by values $-6.17$ eV, $-6.61$ eV, $-6.98$ eV,
  $-7.23$ eV, $-7.36$ eV; and LUMO approaches to $E_f$
  by values $-2.84$ eV, $-3.34$ eV, $-3.86$ eV,
  $-4.37$ eV, and $-4.83$ eV, respectively.

  With the gate bias $V_g=5.00$ V, the separation between $E_{f}$ and LUMO,
  $|E_{f}-\mbox{LUMO}|=0.27$ eV, which leads the large current at small bias.
  The I-V curves of anthracene-dithiol controlled by the gate bias are shown in Fig. 6.
  The current magnitude of the anthracene-dithiol is one order smaller than the one of
  the fused-ring thiophene molecules, which is in favor of the low-power device.
  The molecular transistor
  illustrates the effective current separation under control of
  the positive gate bias. For the applied voltage $-1.0$ V,
  $(\mbox{I}_{V_g=+2.50 V}-\mbox{I}_{V_g=0 V})/(\mbox{I}_{V_g=0 V})\doteq
  0.61$, $(\mbox{I}_{V_g=+5.00 V}-\mbox{I}_{V_g=0 V})/(\mbox{I}_{V_g=0 V})\doteq 5.89$;
  while for $-2.0$ V applied voltage,
  $(\mbox{I}_{V_g=+2.50 V}-\mbox{I}_{V_g=0 V})/(\mbox{I}_{V_g=0 V})\doteq
  3.05$, $(\mbox{I}_{V_g=+5.00 V}-\mbox{I}_{V_g=0 V})/(\mbox{I}_{V_g=0 V})\doteq 9.14$.
  We notice that in the voltage range from $-1.0$ V to $-2.0$ V, the current
  under the transverse field of $V_g=-5.00$ V is bigger than that
  at $V_g=0$ V, which is not found for fused-ring oligothiophene molecules .
  The gate-controlled current separation for the negative bias is not
  effective, we have to use its positive-bias controlled function.
  For the anthracene-dithiol, the transverse field of
  $V_g=-5.00$ V makes HOMO close to $-6.0$ eV, accompanying the
  sudden rise of current around $-2.0$ V voltage, so its current is larger
  than the one for $V_g=0$ where neither HOMO nor LUMO enters the
  voltage window.
  The inset of Fig. 6 illustrates the electron number deviation from equilibrium.

  For the open system without the transverse bias,
  the $\alpha$ electron number has almost
  no variation below the applied voltage $V=2.0$ V, since both HOMO and LUMO keep away
  from $E_{f}$ in the voltage range. With increase of gate bias, LUMO is shifted to $E_{f}$ gradually, and the electrons entering
  the molecule are more than that leaving the molecule, which
  makes the electron number increase.
  With decrease of bias, HOMO is close to $E_{f}$ gradually, which makes the electron number decrease.

\begin{figure}
\includegraphics[scale=0.5,angle=0,bb=15 15 558 807]{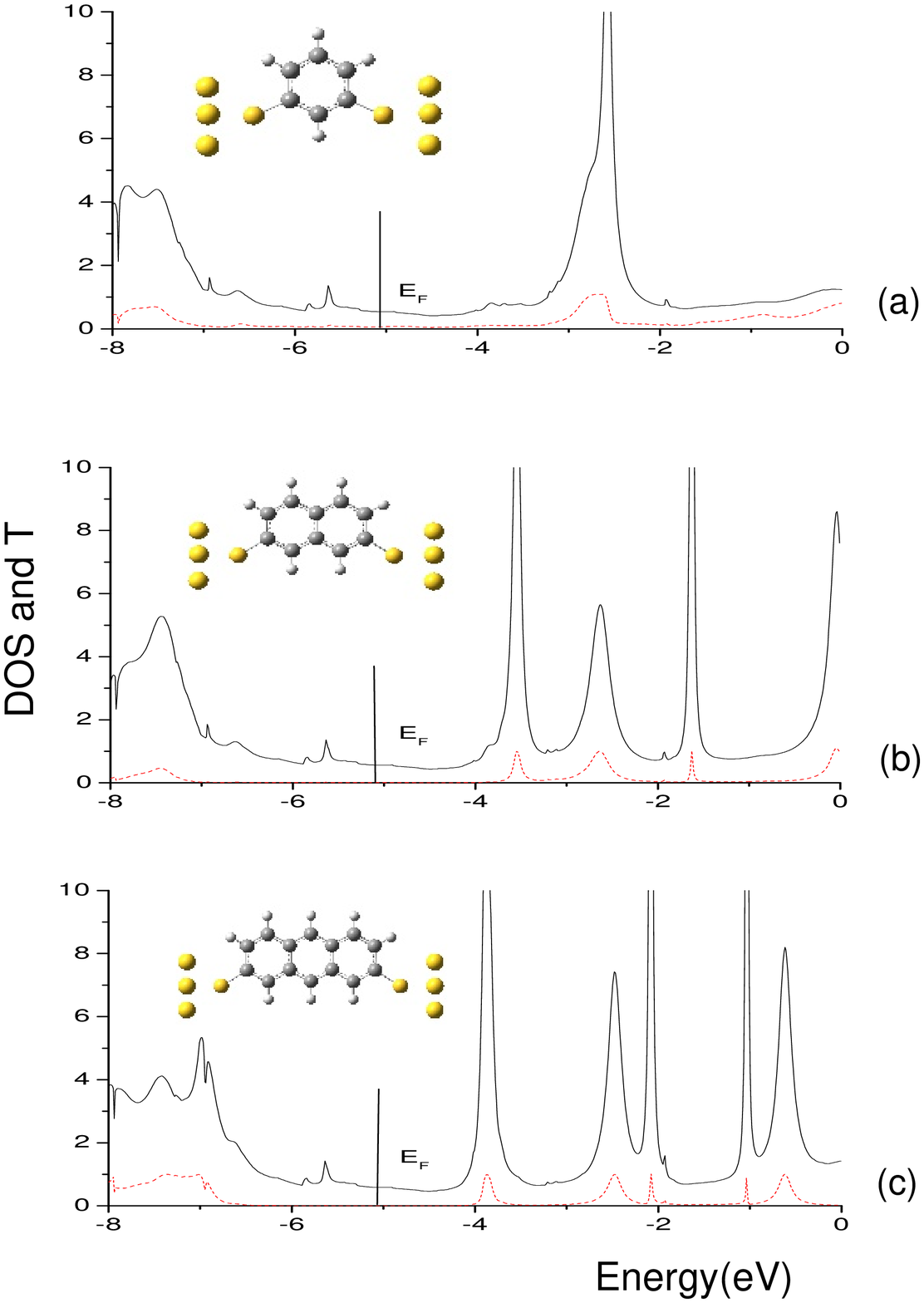}
\caption{DOS (solid) and T (dashed) as functions of energy
corresponding to the different aromatic phenyl ring molecules with
gold contacts for single PDT (a), naphthalene-dithiol (b), and
anthracene-dithiol (c). The vertical line denotes the position of
Fermi level. \label{fig5}}
\end{figure}

  With increase in the number of aromatic phenyl ring, the TFE is
  improved. We found TFE of the naphthalene-dithiol is better
  than that of the single PDT,
  but not better than that of the anthracene-dithiol.
  For the single PDT, under bias $-1.0$ V,
  $(\mbox{I}_{V_g=+2.50 V}-\mbox{I}_{V_g=0 V})/(\mbox{I}_{V_g=0 V})\doteq
  0.03$, $(\mbox{I}_{V_g=+5.00 V}-\mbox{I}_{V_g=0 V})/(\mbox{I}_{V_g=0 V})\doteq 0.11$,
  while under bias -2.0 V, $(\mbox{I}_{V_g=+2.50 V}-\mbox{I}_{V_g=0 V})/
  (\mbox{I}_{V_g=0 V})\doteq 0.04$, $(\mbox{I}_{V_g=+5.00 V}-\mbox{I}_{V_g=0 V})/
  (\mbox{I}_{V_g=0 V})\doteq 0.12$. The control of the molecular current
  by the gate bias is not discernible, which is consistent with the experiment.\cite{31}
  The I-V curves of the pentacene under the different gate bias are shown in Fig. 7.
  Under bias $-1.0$
  V, the molecular device presents the obvious current modulation by the gate-bias:
 $(\mbox{I}_{V_g=+1.25 V}-\mbox{I}_{V_g=0 V})/(\mbox{I}_{V_g=0 V})\doteq 0.32$;
 $(\mbox{I}_{V_g=+1.875 V}-\mbox{I}_{V_g=0 V})/(\mbox{I}_{V_g=0 V})\doteq 1.31$;
 $(\mbox{I}_{V_g=+2.50 V}-\mbox{I}_{V_g=0 V})/(\mbox{I}_{V_g=0 V})\doteq 3.23$.
 The molecule illustrates the five times of the current modulation
 rate of the anthracene-dithiol under the same gate bias.
 Considering the stability of the structure, maybe the
 anthracene-dithiol is still a better candidate for the molecular
 transistor.

\begin{figure}
\includegraphics[scale=0.5,angle=-90,bb=23 38 571 782]{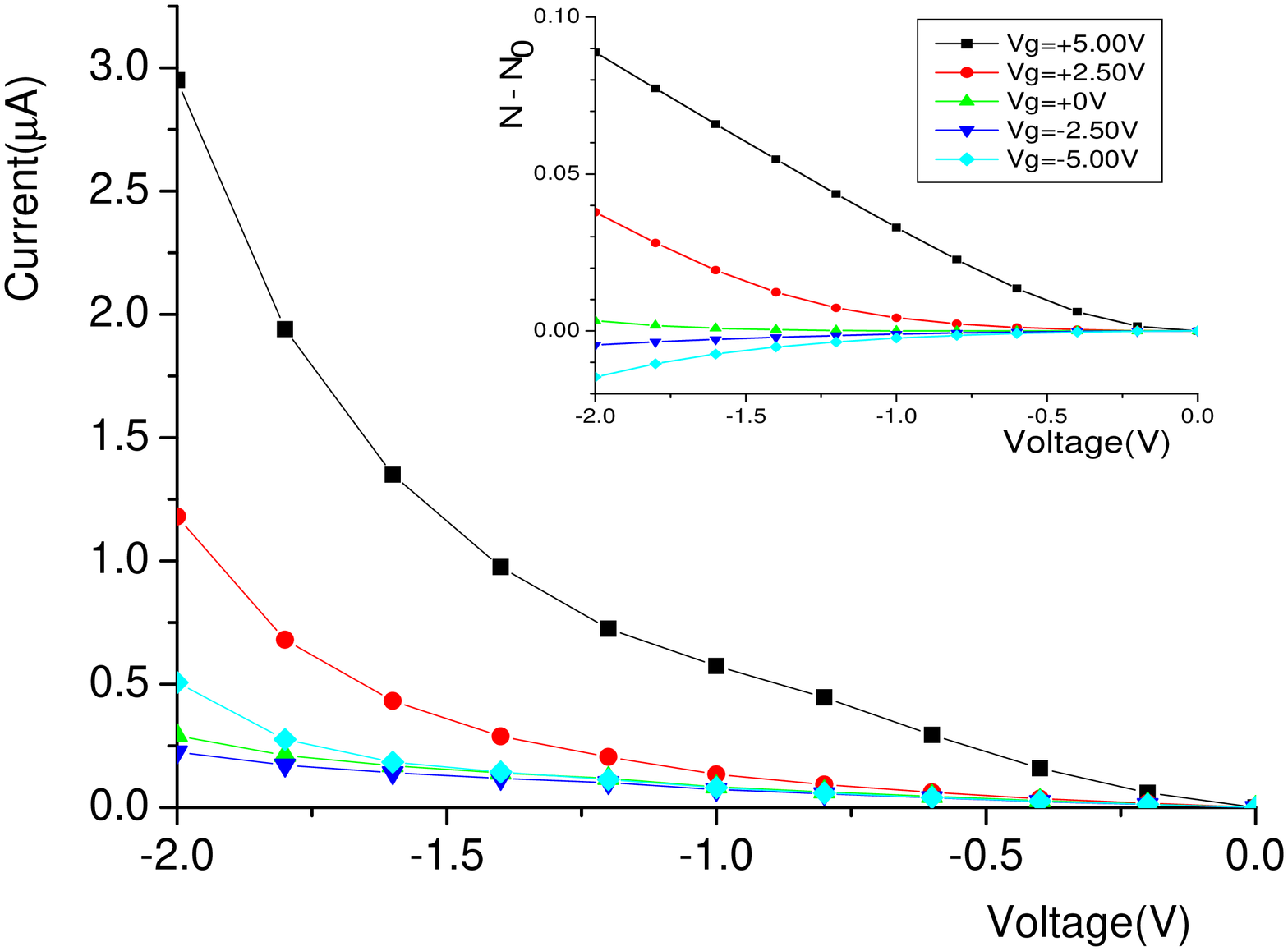}
\caption{I-V curves of pentacene corresponding to the different
gate bias. The inset shows the variation of $\alpha$ electron
number relative to that in equilibrium with the change of
bias.\label{fig6}}
\end{figure}

\begin{figure}
\includegraphics[scale=0.5,angle=-90,bb=34 56 577 755]{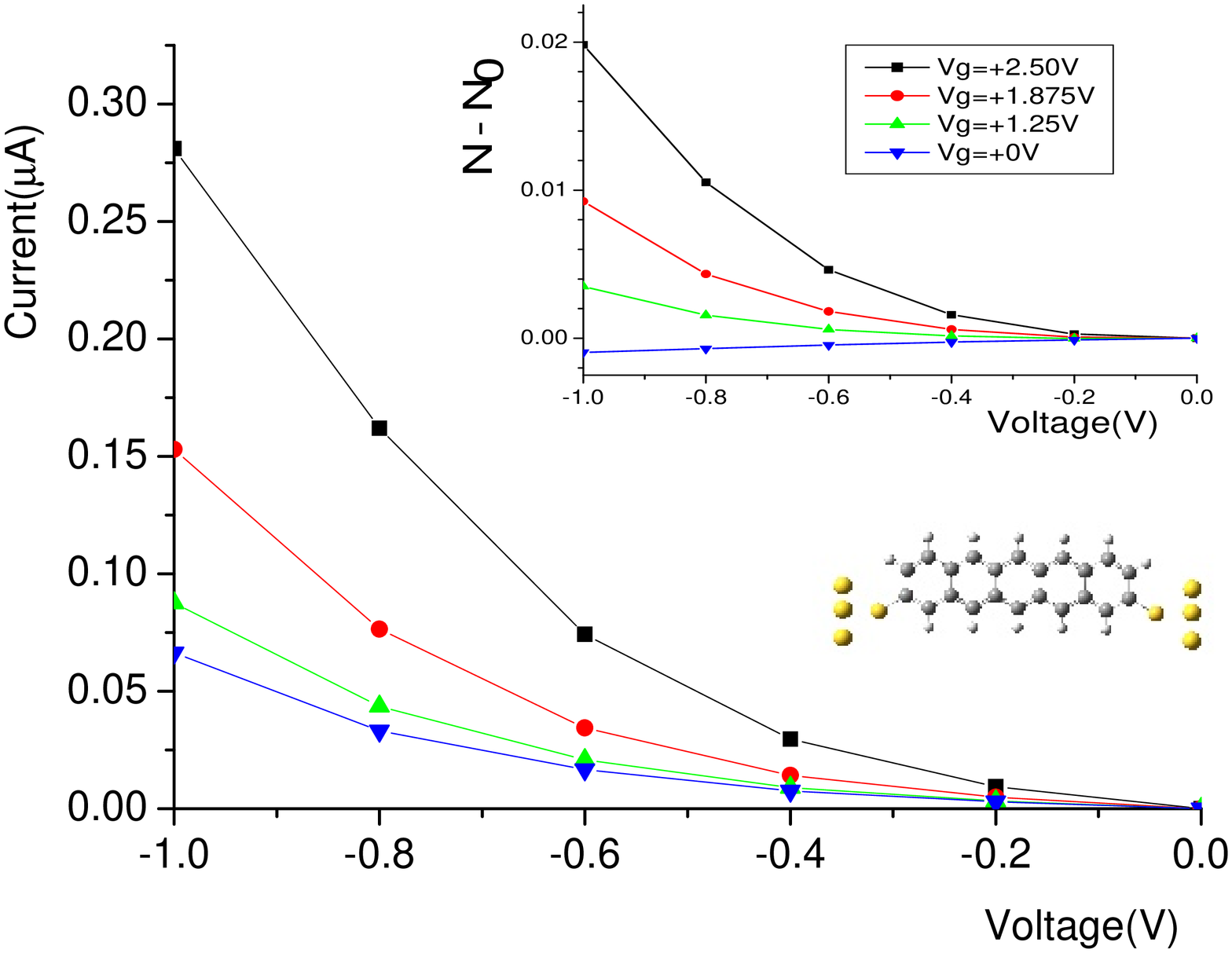}
\caption{I-V curves of anthracene-dithiol corresponding to the
different gate bias. The inset shows the variation of $\alpha$
electron number relative to that in equilibrium with the change of
bias.\label{fig7}}
\end{figure}

The above TFE comes from the fact that with the increase in
transverse field for the polyacene-dithiol, it will make energy
levels lower and for the fused ring thiophene, with the increase
in the transverse field, it will make energy levels higher. It
should be noted that the energy level shift is determined by the
choose of the potential zero point, which is set at the center of
the line connected by two sulfur atoms in our model. If the
potential zero point deviates from this point, the expression of
chemical potential of left and right leads should be rectified:
$\mu_{1}=\mbox{$E_{f}$}-e(\frac{1}{2}V+\Delta V)$, and
$\mu_{2}=\mbox{$E_{f}$}+e(\frac{1}{2}V-\Delta V)$. $\Delta V$ is
the electrical potential difference between the old zero point and
the new one. However, the choose of zero point has no effect on
the calculation results.

\section{\label{sec:level1}SUMMARY}
We use the self-consistent method based on the DFT and
non-equilibrium Green's function to simulate molecular transport.
In terms of Gaussian03, the electronic structures of the molecular
device and the macroscopic leads are calculated on an equal
footing. At the same time, the self-consistent iteration cycle is
extended from the local molecule to the open system, by inserting
the calculation of the density matrix of the open system as a
subroutine. Our self-consistent results show that the long-length
organic molecule can achieve better transport characteristics. Our
investigation proves that both the polyacene-dithiol molecules and
fused-ring oligothiophene molecules can be made as the
high-frequency molecular transistors controlled by the transverse
field. The theoretical results suggest the related experiments
about the molecular devices.

\begin{acknowledgments}
This work is supported by the National Natural Science Foundation
of China (NSFC) under Projects 90206031 and 10574024. The authors
gratefully acknowledge SR8000 supercomputer resources from the
Center for Computational Materials Science of the Institute for
Materials Research, Tohoku University.
\end{acknowledgments}

\end{document}